\documentclass[aps,prb, superscriptaddress,
 amsmath,amssymb,
 reprint,%
]{revtex4-2}

\usepackage{graphicx}
\usepackage{subfigure}
\usepackage{dcolumn}
\usepackage{multirow}
\usepackage{bm}
\usepackage{array}
\usepackage[colorlinks=true,linkcolor=blue,anchorcolor=red,citecolor=blue,urlcolor=black]{hyperref}
\usepackage{comment}
\usepackage[table,xcdraw]{xcolor}
\usepackage{float}
\usepackage{longtable}
\usepackage{mathptmx}
\usepackage{amsmath}
\usepackage{etoolbox, xcolor}
\usepackage{soul}
\usepackage{amssymb}
\usepackage{pifont}

\makeatletter
\def\ignorecitefornumbering#1{%
     \begingroup
         \@fileswfalse
         #1
    \endgroup
}
\makeatother

\begin{document}

\title{Ultrathick MA$_2$N$_4$(M'N) Intercalated Monolayers with \\Sublayer-Protected Fermi Surface Conduction States: Interconnect and Metal Contact Applications}

\author{Che Chen Tho}
\thanks{These authors contributed equally.}
\affiliation{ 
Science, Mathematics and Technology Cluster, Singapore University of Technology and Design, Singapore 487372}%

\author{Xukun Feng}
\thanks{These authors contributed equally.}
\affiliation{ 
Science, Mathematics and Technology Cluster, Singapore University of Technology and Design, Singapore 487372}%

\author{Zhuoling Jiang}
\affiliation{Science, Mathematics and Technology Cluster, Singapore University of Technology and Design, Singapore 487372}

\author{Liemao Cao}
\affiliation{College of Physics and Electronic Engineering, Hengyang Normal University, Hengyang 421002, China}

\author{Guangzhao Wang}
\affiliation{Key Laboratory of Extraordinary Bond Engineering and Advanced Materials Technology of Chongqing, School of Electronic Information Engineering, Yangtze Normal University, Chongqing 408100, People’s Republic of China}

\author{Chit Siong Lau}
\affiliation{Institute of Materials Research and Engineering (IMRE), Agency for Science, Technology and Research (A*STAR), Singapore 138634}

\author{San-Dong Guo}
\email{sandongguo@sutd.edu.sg}
\affiliation{School of Electronic Engineering, Xi’an University of Posts and Telecommunications, Xi’an 710121, China}

\author{Yee Sin Ang}
\email{yeesin\_ang@sutd.edu.sg}
\affiliation{ 
Science, Mathematics and Technology Cluster, Singapore University of Technology and Design, Singapore 487372}%

\begin{abstract}

Recent discovery of ultrathick $\mathrm{MoSi_2N_4(MoN)_n}$ monolayers open up an exciting platform to engineer 2D material properties via intercalation architecture. Here we computationally investigate a series of ultrathick MA$_2$N$_4$(M'N) monolayers (M, M' = Mo, W; A = Si, Ge) under both \emph{homolayer} and \emph{heterolayer} intercalation architectures in which the same and different species of transition metal nitride inner core layers are intercalated by outer passivating nitride sublayers, respectively.
The MA$_2$N$_4$(M'N) monolayers are thermally, dynamically and mechanically stable with excellent mechanical strength and metallic properties.
Intriguingly, the metallic states around Fermi level are localized within the inner core layers. Carrier conduction mediated by electronic states around the Fermi level is thus spatially insulated from the external environment by the \emph{native} outer nitride sublayers, suggesting the potential of MA$_2$N$_4$(M'N) in back-end-of-line (BEOL) metal interconnect applications.
Nitrogen vacancy defect at the outer sublayers creates `punch through' states around the Fermi level that bridges the carrier conduction in the inner core layers and the outer environment, forming a electrical contact akin to the `vias' structures of metal interconnects.
We further show that MoSi$_2$N$_4$(MoN) can serve as a quasi-Ohmic contact to 2D WSe$_2$.
These findings reveal the promising potential of ultrathick MA$_2$N$_4$(MN) monolayers as metal electrodes and BEOL interconnect applications. 

\end{abstract}

\maketitle

\section{\label{sec:introduction}Introduction}

Two-dimensional (2D) materials provide an avenue to revolutionize the solid-state device technology beyond the reach of conventional bulk materials, ranging from conventional computing electronics and renewable energy \cite{Quhe2021,Tan2020,Wang2022_catalyst,Song2018} to emerging technology such as spintronics \cite{Ahn_2020, Zhang2021}, valleytronics \cite{Goh2023_book, Vitale2018, Schaibley2016, val1, val2} and neuromorphic devices \cite{huh2020memristors}. 
Recent discovery of septuple-layered semiconducting monolayers of MoSi$_2$N$_4$ and WSi$_2$N$_4$, where an unstable transition metal dinitride layer (e.g. MoN$_2$) is sandwiched and stabilized by two SiN outer layers \cite{Hong2020}, and the broader family of MA$_2$Z$_4$ (M = transition metals; A = Group IVA elements; Z = Group VA elements) family \cite{Wang2021_MA2Z4} reveals another exciting material platform to explore both fundamental physics and device applications unique to 2D materials \cite{Yin2022, Tho2023_review, wozniak2023electronic}. Myriads of potential applications, such as transistor \cite{Zhao2021_transistor, Wang2021, Cao2021, Qu2023}, solar cell \cite{Tho2022, Guo2022_BP}, sensors \cite{xiao2022_gas}, spin transistor \cite{Ren2022}, based on MA$_2$Z$_4$ and their heterostructures have been proposed recently.

The intercalation morphology of MA$_2$Z$_4$ can be expanded to create ultrathick MoSi$_2$N$_4$(MoN)$_{n}$ ($n = 1, 2, 3, \cdots$) monolayers \cite{NSR}. Recent experiment has demonstrated ultrathick MoSi$_2$N$_4$(MoN)$_{4}$ and MoSi$_2$N$_4$(MoN)$_{8}$ \emph{metallic} monolayers formed \emph{natively} between semiconducting MoSi$_2$N$_4$ monolayers \cite{NSR}. 
Such natively-grown metallic layered materials with layered semiconductors represents a metallic counterpart to the natively-grown layered oxides on layered semiconductor, such as Bi$_2$SeO$_5$/Bi$_2$O$_2$Se \cite{li2020native, zhang2022single} and Bi$_2$TeO$_6$/Bi$_2$O$_2$Te \cite{zou20232d}. 
The inner core layer number ($n$) and hence the thickness of a single MoSi$_2$N$_4$(MoN)$_n$ monolayer, can be modulated by growth parameters. 
Such tunable morphology reveals an avenue for controlling the thickness of a single monolayer via \emph{sublayer number engineering} approach. 

\begin{figure*}[t]
\includegraphics[width=0.9\textwidth]{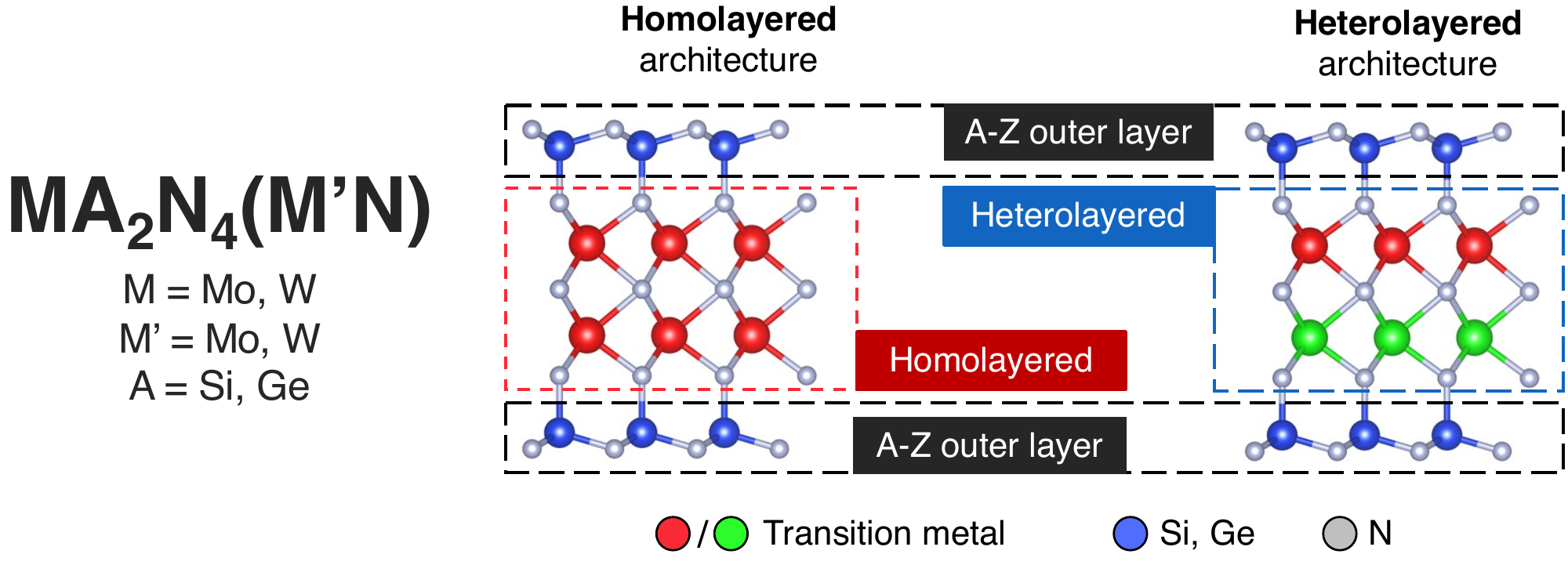}
\caption{\label{Fig1}\textbf{Morphology of ultrathick MA$_2$Z$_4$(M'Z) in \emph{homolayer} and \emph{heterolayer} morphologies.} MA$_2$Z$_4$(M'Z) monolayers constructed with the intercalation of $\alpha$-A$_2$Z$_2$ with 2H-MZ$_2$ sublayers. }
\end{figure*}

The experimental synthesis of MoSi$_2$N$_4$(MoN)$_n$ suggests the existence of a broader MA$_2$Z$_4$(MZ)$_n$ family. MA$_2$Z$_4$(MZ)$_n$ is a \emph{homologous} layered material \cite{Mrotzek2003, Kanatzidis2005} constructed using MA$_2$Z$_4$ as a blueprint. The inner MZ$_2$ sublayers are expanded along the out-of-plane direction \cite{NSR} by stacking multiple copies of MZ sublayers between the two AZ$_2$ outer sublayers. 
\emph{Homolayer} intercalation of MZ$_2$(MZ)$_n$ is obtained by inserting $n$ copies of identical MZ innerlayer between the outer AZ sublalyers. Preliminary computational studies accompanying the experimental observation of MoSi$_2$N$_4$(MoN)$_n$ reveals their structural stability and a metallic nature for $n$ up to 4 \cite{NSR}. 
Beyond the \emph{homolayer} architecture, \emph{heterolayer} intercalation, where the inner layers are composed of sublayers with different transition metal atoms (i.e. sublayers of MN and M'N where M$\neq$M'), can also be constructed (see Figure \ref{Fig1} for an illustration of \emph{homolayered} and \emph{heterolayered} monolayers with $n=1$) and may host richer physical properties due to the broken mirror symmetry along the out-of-plane direction.

Motivated by the recent experimental synthesis of MoSi$_2$N$_4$(MoN)$_4$, the goals of this work are: (i) to provide a zoom-in understanding on the physical properties of MA$_2$N$_4$(M'N) where M, M' = Mo, W; and A = Si, Ge; (ii) to provide a pilot study on \emph{heterolayered} monolayers where M$\neq$M'; and (iii) to understand the device application potentials of MA$_2$N$_4$(M'N). 
Using first-principles density functional theory (DFT) simulations, we examine six species of MA$_2$N$_4$(M'N) monolayers (see Figure \ref{Fig1}) under both \emph{homolayer} and \emph{heterolayer} architectures. 
MA$_2$N$_4$(M'N) are dynamically, thermally and mechanically stable. 
Such monolayers exhibit excellent mechanical strength comparable to MoSi$_2$N$_4$ \cite{Hong2020}, and exhibit metallic properties. 
For \emph{heterolayered} MA$_2$N$_4$(M'N), the broken out-of-plane inversion symmetry leads to a built-in electric field and induces a Rashba spin-orbit coupling effect. 

Intriguingly, the electronic states around the Fermi level, which are responsible for carrier conduction, are spatially distributed within the inner MN and M'N sublayers, thus unraveling an intrinsic \emph{sublayer protection} mechanism in which the outer AN nitride sublayers protect the conduction states lying in the MN core layers -- a behavior akin to the sublayer protection of the band edge state in MoSi$_2$N$_4$ and WSi$_2$N$_4$ \cite{Wang2021, wu2022prediction}. 
Using MoSi$_2$N$_4$(MoN) as an illustrative example, we show that nitrogen vacancy defect can effectively create a conduction pathway that enables the inner core conduction states to `punch through' the inert SiN outer layers, forming a \emph{via} connection to the outer environment. 
Such a behavior suggests the potential application of MoSi$_2$N$_4$(MoN) as a sub-1-nm thick metal interconnect -- a critical component of back-end-of-line (BEOL) applications \cite{lo2020opportunities} -- that are robust against environmental influences and the electrical \emph{vias} contact to the front-end-of-line (FEOL) transistors can be created through defect engineering.
We further show that the metal/semiconductor contact heterostructure composed of MoSi$_2$N$_4$(MoN) and WSe$_2$ exhibits an $n$-type ultralow Schottky barrier height of 25 meV, which is beneficial for constructing energy-efficient metal contact to 2D semiconductor field-effect transistor \cite{infomat}.  
These findings reveal MA$_2$N$_4$(M'N) as a compelling metallic platform for interconnect and metal contact applications, and may provide a key material component for BEOL in the next-generation integrated circuit technology.

\begin{table*}[]

\caption{\label{Elastic} Optimised lattice constants, monolayer thickness ($d$), elastic constants ($C_{11}$, $C_{12}$, $C_{66}$), 2D and 3D Young's modulus ($Y^{2D}$, $Y^{3D}$), 3D bulk modulus ($K^{3D}$), 3D sheer modulus ($G^{3D}$) and Poisson's ratio ($\nu$) of the 2D materials.} 

\resizebox{\textwidth}{!}{%

\begin{tabular}{>{\centering\arraybackslash}m{2.5cm}>{\centering\arraybackslash}m{2cm}>{\centering\arraybackslash}m{2cm}>{\centering\arraybackslash}m{1.5cm}>{\centering\arraybackslash}m{1.5cm}>{\centering\arraybackslash}m{1.5cm}>{\centering\arraybackslash}m{2.5cm}>{\centering\arraybackslash}m{2cm}>{\centering\arraybackslash}m{1.5cm}>{\centering\arraybackslash}m{1.5cm}>{\centering\arraybackslash}m{1.5cm}}

\hline \hline  
\textbf{Materials} & \textbf{Lattice   Constant ($\AA$)} & \textbf{d ($\AA$)} & \textbf{C$_{11}$ (N/m)} & \textbf{C$_{12}$ (N/m)} & \textbf{C$_{66}$ (N/m)} & \textbf{$Y^{2D}$ (N/m)} & \textbf{$Y^{3D}$ (GPa)} & \textbf{$K^{3D}$ (GPa)} & \textbf{$G^{3D}$ (GPa)} & \textbf{$\nu$} \\ \hline
MoSi$_2$N$_4$(MoN)       & 2.92                        & 9.81               & 663                & 161                & 251                & 624                                                                      & 484                              & 319                           & 195                            & 0.24                     \\ 
MoSi$_2$N$_4$(WN)        & 2.92                        & 9.83               & 685                & 145                & 270                & 654                                                                      & 506                              & 321                           & 209                            & 0.21                     \\ 
WSi$_2$N$_4$(WN)         & 2.93                        & 9.85               & 697                & 142                & 277                & 668                                                                      & 516                              & 324                           & 214                            & 0.20                     \\ 
MoGe$_2$N$_4$(MoN)       & 3.02                        & 10.28              & 580                & 178                & 201                & 525                                                                      & 392                              & 283                           & 150                            & 0.31                     \\
MoGe$_2$N$_4$(WN)        & 3.03                        & 10.3               & 585                & 176                & 204                & 532                                                                      & 397                              & 283                           & 153                            & 0.30                     \\ 
WGe$_2$N$_4$(WN)         & 3.03                        & 10.33              & 582                & 181                & 201                & 526                                                                      & 391                              & 284                           & 149                            & 0.31                     \\ 

MoSi$_2$N$_4$(MoN)/WSe$_2$ & 5.83 & 16.67 & 790 & 181 & 305 & 749 & 372 &  241 & 151 & 0.23 \\ \hline \hline
\end{tabular}%
}
\end{table*}

\section{\label{sec:Methods}Computational Methods}

We perform first-principles calculations from density functional theory (DFT) using the Vienna Ab initio Simulation Package (VASP) \cite{Hafner2008}. To eliminate spurious interaction from the periodic images along the out-of-plane direction, a vaccuum thickness of 20 $\mathrm{\AA}$ is constructed at each end of the lattice structure and dipole correction is taken into account. The Perdew-Burke-Ernzerhof generalised gradient approximation (PBE-GGA) was chosen for the exchange-correlation functional in our structural optimization and electronic properties calculations. Since all the studied materials are metallic, the choice of either PBE-GGA or Heyd-Scuseria-Ernzerhof hybrid form (HSE06-GGA) will not cause much difference to the obtained results. A Brillouin zone k-point sampling grid of 6 $\times$ 6 $\times$ 1 was used for structural relaxation. For self-consistent field calculations, a denser $\Gamma$-centered Brillouin zone k-point sampling grid of 12 $\times$ 12 $\times$ 1 was used. The break condition for each ionic relaxation loop is set to 10$^{-3}$ eV Å$^{-1}$ for the forces acting on the unit cell, and the break condition for each electronic relaxation loop is set to 10$^{-6}$ eV. We adopted the optB88-vdW functional for van der Waals (vdW) correction. QuantumATK \cite{Quantumatk} was used to construct MoSi$_2$N$_4$(MoN)/WSe$_2$ vdWH with a lattice mismatch of 2.47\% between 2 $\times$ 2 unit cells of MoSi$_2$N$_4$(MoN) and $\sqrt{3} \times \sqrt{3}$ unit cells of WSe$_2$ with the appropriate rotation.

\section{\label{sec:Results}Results and Discussions}

\subsection{\label{Structure and Stability} Structural and mechanical properties, and stability of MA$_2$Z$_4$(M'Z)}

\begin{figure*}
\centering
\includegraphics[width=0.958\textwidth]{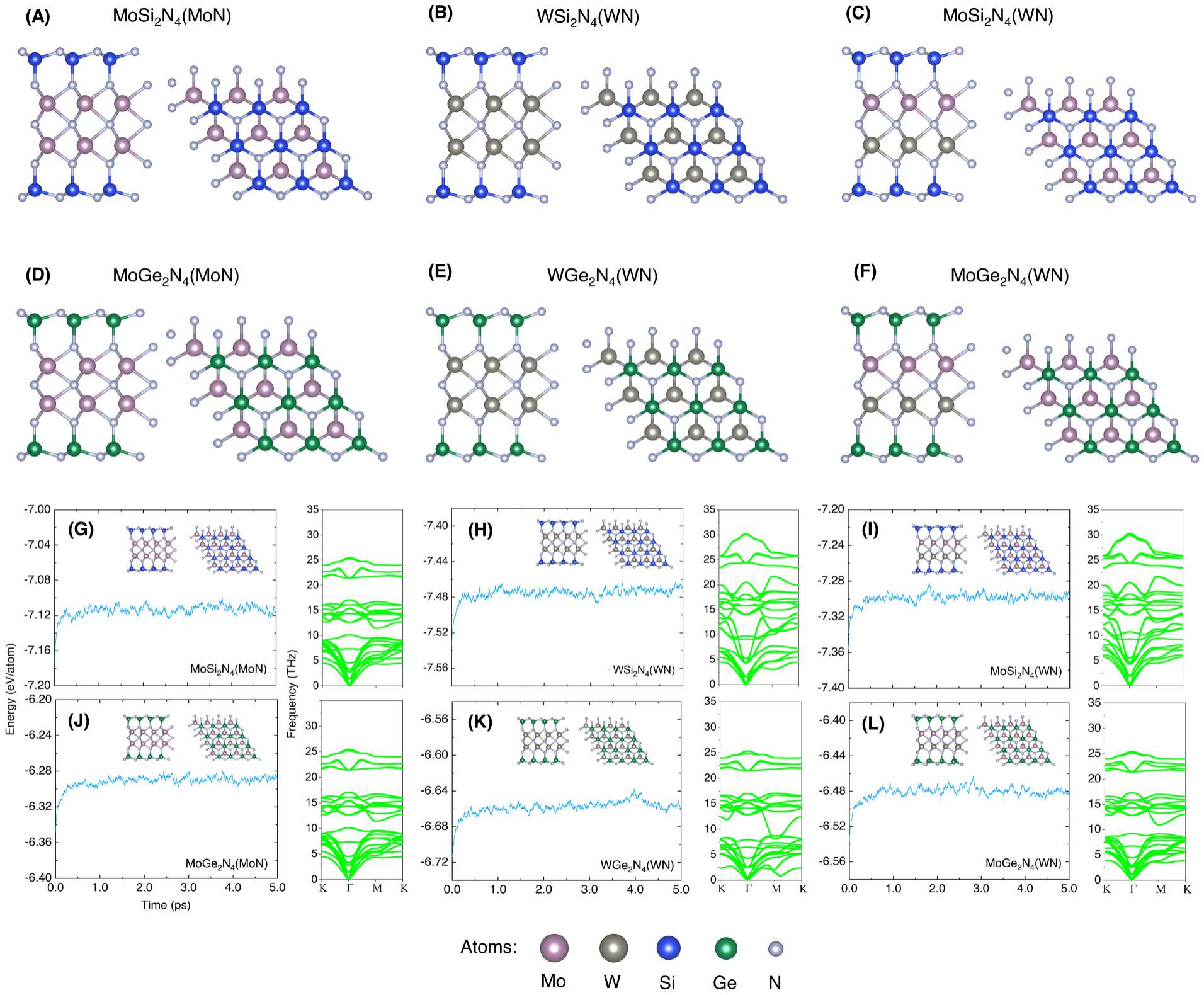}
\caption{\label{Fig2}\textbf{Lattice structure and stability tests.} (A-F) Top and side views of the monolayer lattice structures. (G-L) AIMD plots and phonon spectrums showing the dynamical and thermodynamical stabiility of the monolayers.}
\end{figure*}

\textbf{Structural Properties.} 
We investigate six representative MA$_2$N$_4$(M'N) monolayers. The side and top views of the lattice structures are shown in Figure \ref{Fig2} (A-F). The monolayers consist of nine atoms, in the order of N-A-N-M-N-M-N-A-N along the out-of-plane direction, where the sandwiched MN$_2$(MN) sublayer is a composed of two overlapping 2H-MN$_2$ sublayers sandwiched between two outer $\alpha$-AN sublayers. The MoSi$_2$N$_4$(MoN), WSi$_2$N$_4$(WN), MoGe$_2$N$_4$(MoN) and WGe$_2$N$_4$(WN) have an \emph{homolayer} intercalation morphology with the space group 187 (P6m2) and the point group D$_{3h}$. The C$_3$ principal axis of rotation passes through the inner three N and the Si (Ge) atoms for these monolayers. Out-of-plane mirror symmetry of these structures implies a $\sigma_h$ plane that passes through the center N atom, and there are three C$_2$ axes at the center N atom that are perpendicular to the C$_3$ principal axis separated at 120$^\circ$ apart. The \emph{heterolayered} monolayers counterpart, i.e. Mo(Si,Ge)$_2$N$_4$(WN), on the other hand have an \emph{heterolayer} intercalation structure, with the absence of a mirror symmetry and the three C$_2$ axes. These \emph{heterolayered} monolayers belong to the space group 156 (P3m1) and point group C$_{3v}$. The lattice constants of the six monolayers after geometrical optimization are given in Table \ref{Elastic}. The lattice constants and monolayer thickness of MoSi$_2$N$_4$(MoN) calculated in this work are in agreement with those reported in a previous work \cite{NSR}.

\begin{figure*}[t]
\centering
\includegraphics[width=0.85\textwidth]{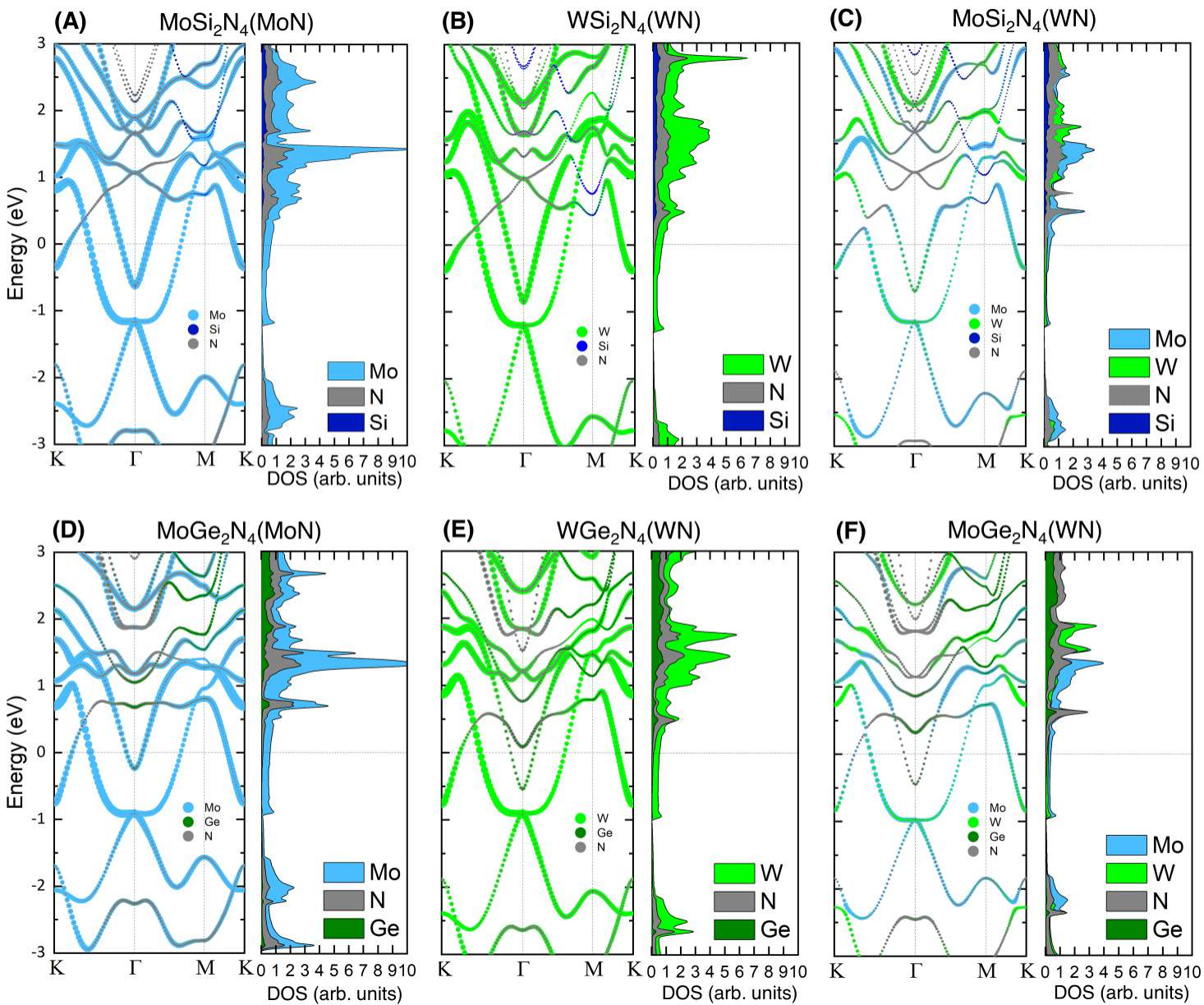}
\caption{\label{Fig3} \textbf{Electronic band structures and density of states} Non-spin polarised calculations of the electronic band structures reveal that all monolayers are metallic. The energy states around the Fermi level (E$\mathrm{_F}$) are mainly contributed by the Mo (W) atoms and traces from N atoms, whereas Si (Ge) atoms have the least contribution throughout the displayed energy range.}
\end{figure*}

\textbf{Mechanical properties.} The elastic constants (C$_{i,j}$) is obtained using the energy-strain method under the harmonic approximation implemented in VASPKIT \cite{Wang2021_vaspkit}:
\begin{equation}\label{Energy_Strain}
\Delta U = \frac{V_0}{2}\sum_{i=1}^{6}\sum_{j=1}^{6}C_{i,j}e_{i}e_{j}
\end{equation}
where $\Delta U$ is the energy difference between the distorted and undistorted lattice cell, $V_0$ is the volume of the undistorted lattice cell, and e$_i$(e$_j$) is the matrix element of the strain vector. For 2D isotropic monolayers with in-plane hexagonal symmetry like graphene, there are only three elastic constants ($C_{11}$, $C_{12}$, $C_{66}$ $= (C_{11}-C_{12})/2$) that determines the material's mechanical properties. The 2D Young's modulus ($Y^{2D}$) and the Poisson's ratio ($\nu$) are calculated as $Y^{2D} = (C_{11}^2-C_{12}^2)/C_{11}$ and $\nu = C_{12}/C_{11}$, respectively. In order to make a comparison across materials of different thickness, values of $Y^{2D}$ are normalized with respect to the thickness ($d$) of the monolayers to obtain 3D Young's Modulus ($Y^{3D}$), using $Y^{3D} = Y^{2D}/(d + 2d_{N, vdW})$ where $d_{N, vdW} = 1.55$ \AA is the vdW radius of N atom. The 3D bulk modulus ($K^{3D}$) and 3D shear modulus ($G^{3D}$) are calculated as $K^{3D} = Y^{3D}/(2-2\nu)$ and $G^{3D} = Y^{3D}/(2+2\nu)$, respectively [see Table \ref{Elastic} for elastic constants and various moduli]. The $Y^{3D}$ of MoSi$_2$N$_4$(MoN), WSi$_2$N$_4$(WN), MoSi$_2$N$_4$(WN) are close to that of the MoSi$_2$N$_4$ (493 GPa) \cite{Hong2020}, which are higher than several commonly studied 2D materials (e.g. TMDC: MoS$_2$ \cite{Cooper2013,Liu2014}, WS$_2$ \cite{Liu2014}; MXenes: Ti$_3$C$_2$T$_x$ \cite{Lipatov2018}, Nb$_4$C$_3$T$_x$ \cite{Lipatov2020}).

\begin{figure*}[t]
\centering
\includegraphics[width=0.7\textwidth]{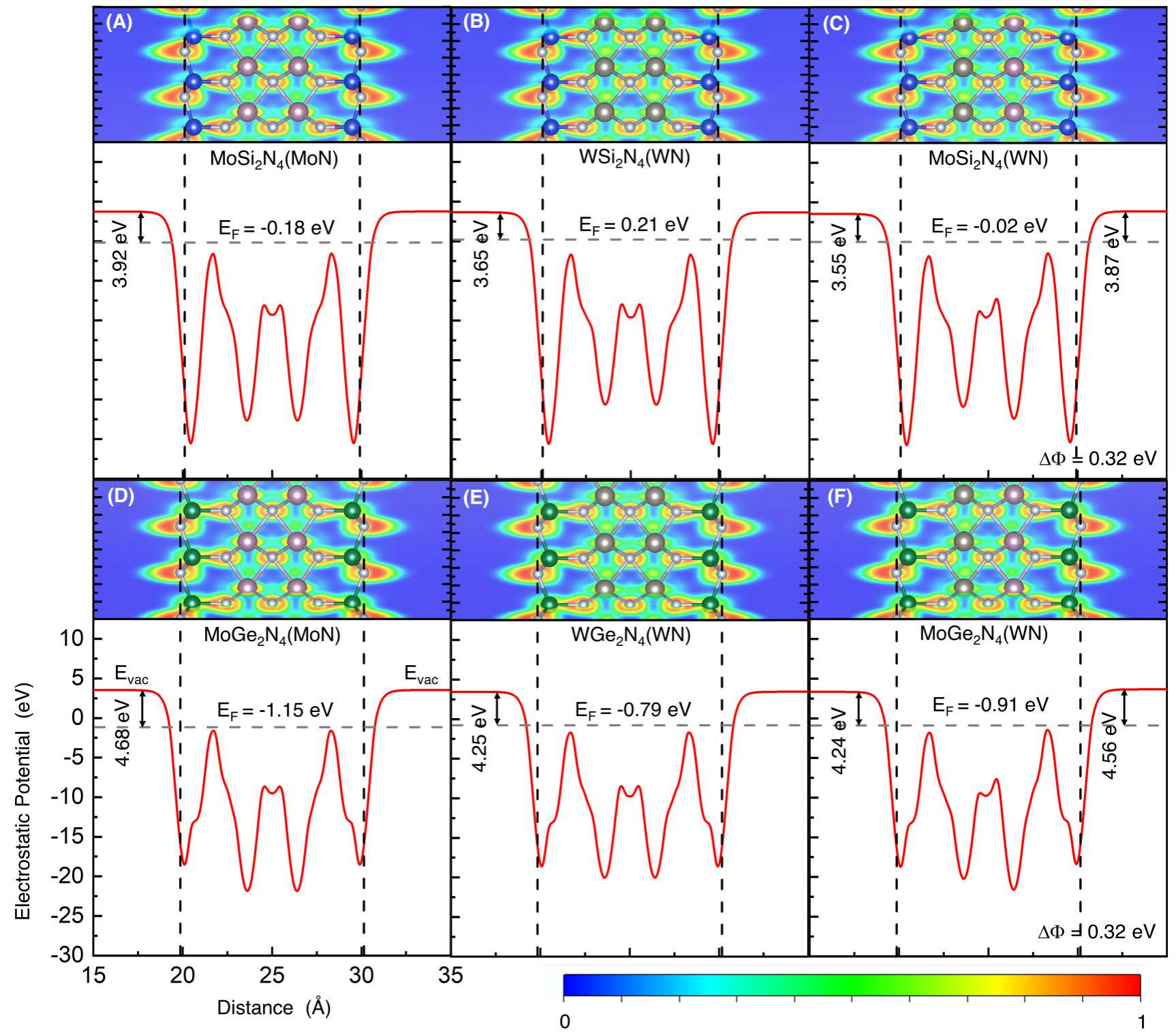}
\caption{\label{Fig4} \textbf{Electrostatic potential electron localization function profiles.} (A-F) The plane-averaged electrostatic potential profile (bottom panels) and electron localization function (top panels) of the monolayers. E$\mathrm{_{vac}}$, E$\mathrm{_F}$, and $\mathrm{\Delta \Phi}$ denotes the energy of the vacuum level, Fermi level, vacuum offset, respectively. Colour bar describes the normalized valence electron density with red(blue) denoting full localization(delocalization).}
\end{figure*}

\textbf{Stability.} The elastic constants of all MA$_2$N$_4$(MN) monolayers meet the Born-Huang’s requirement, thus confirming their mechanical stability (i.e. $C_{11}$ $>$ $C_{12}$ and $C_{66}$ $>$ 0). Thermodynamic stability of the monolayers are verified using Ab initio molecular dynamics simulation (AIMD). We choose the Andersen thermostat which couples the monolayer to a heat bath at a temperature of 300 K for 5 ps, and set the Anderson probability of collision at 0.5. The atoms are only slightly shifted from their equilibriuim positions with no drastic changes during the period of collisions with random particles of the heat bath, thus confirming their thermodynamical stability. Dynamical stability of the monolayers is verified using PHONOPY program \cite{Phonopy_JPCM, Phonopy_JPSJ} under the finite difference method simulated on a supercell composed of 4 $\times$ 4 unit cells. The absence of soft modes in the phonon spectra confirms the dynamical stability of the monolayers. The free energy versus time fluctuation of the lattice structures as calculated by AIMD, and their phonon spectra calculated by PHONOPY are shown in Figure \ref{Fig1}(G-L). 

\begin{table*}[t]
\centering
\caption{\label{Electronic} Number of electrons (e$^-$) gain/loss by each atom species, the work function of the monolayers W$_M$ (eV) and energy value of the vacuum offset $\Delta \Phi$ (eV) at both ends of the monolayers. "Center", "Inner" and "Outer" denotes N atoms that are located at the center, close to the center, and at the surface, respectively. "/" denotes differing values at each half of the \emph{heterolayered} monolayers.} 
\begin{tabular}{>{\centering\arraybackslash}m{2.5cm}>{\centering\arraybackslash}m{2cm}>{\centering\arraybackslash}m{2cm}>{\centering\arraybackslash}m{1.5cm}>{\centering\arraybackslash}m{1.5cm}>{\centering\arraybackslash}m{1.5cm}>{\centering\arraybackslash}m{2.5cm}>{\centering\arraybackslash}m{2cm}}
\hline \hline
\textbf{Materials} & \textbf{Mo/W} & \textbf{Si/Ge} & \textbf{N (Center)} & \textbf{N (Inner)} & \textbf{N (Outer)} & \textbf{W$_M$} & $\Delta \Phi$   \\ \hline
MoSi$_2$N$_4$(MoN)       & -1.5          & -3             & 1.3                 & 1.6                & 2.2                & 3.92        & 0.00 \\ 
MoSi$_2$N$_4$(WN)        & -1.4/-1.6     & -3             & 1.4                 & 1.6                & 2.2                & 3.87/3.55   & 0.32 \\ 
WSi$_2$N$_4$(WN)         & -1.6          & -3             & 1.4                 & 1.6                & 2.2                & 3.65        & 0.00 \\ 
MoGe$_2$N$_4$(MoN)       & -1.5          & -1.9           & 1.3                 & 1.3                & 1.4                & 4.68        & 0.00 \\ 
MoGe$_2$N$_4$(WN)        & -1.5/-1.7     & -1.9           & 1.4                 & 1.3/1.4            & 1.4                & 4.56/4.24   & 0.32 \\ 
WGe$_2$N$_4$(WN)         & -1.6          & -1.9           & 1.4                 & 1.4                & 1.4                & 4.25        & 0.00 \\ \hline \hline
\end{tabular}%
\end{table*}

\subsection{\label{electronic_monolayer}Electronic and magnetic properties of MA$_2$Z$_4$(MZ)}

\begin{figure*}
\centering
\includegraphics[width=0.958\textwidth]{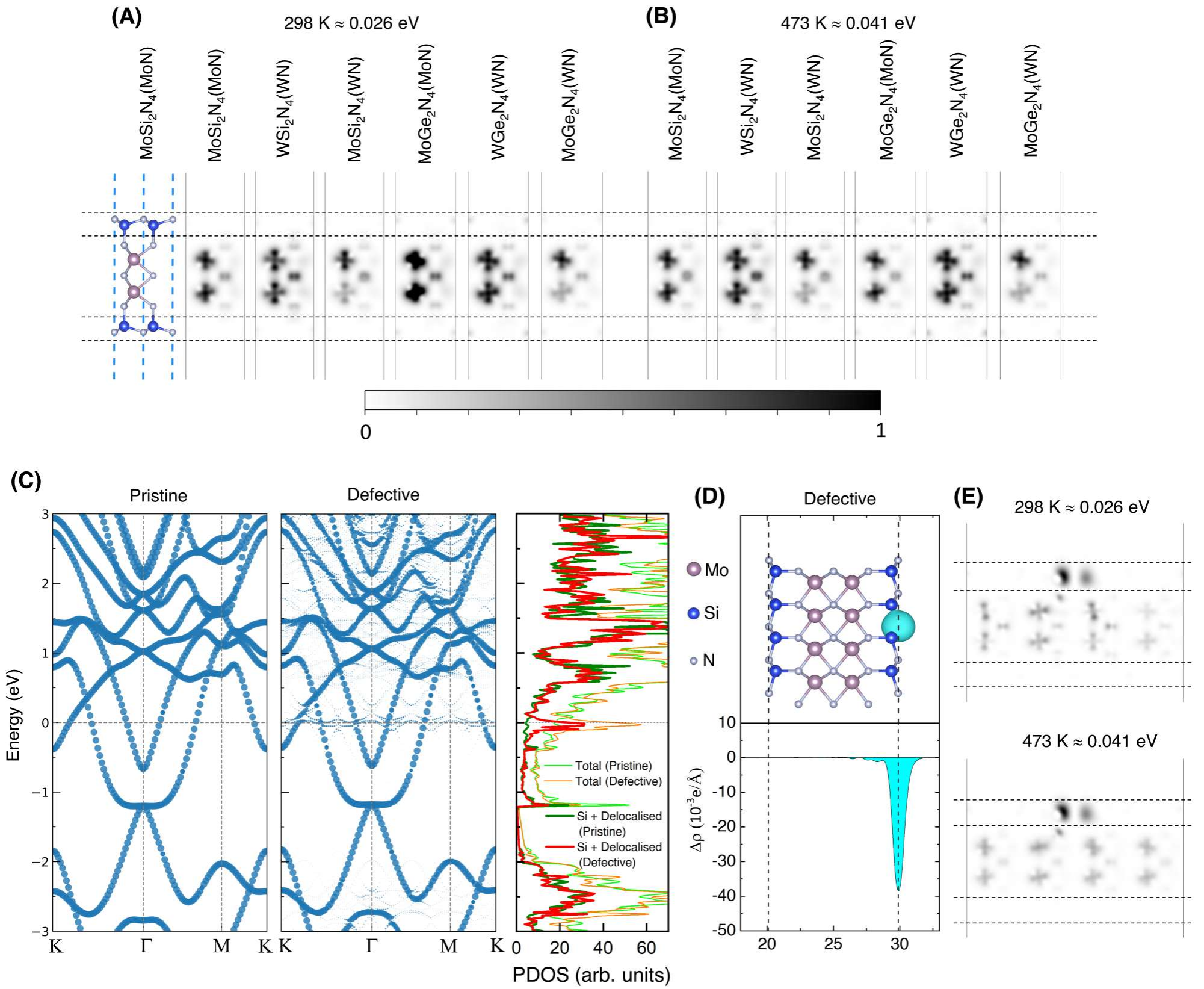}
\caption{\label{Fig5} \textbf{Band decomposed charged density and electronic states of defects.} (A) Band decomposed charge density of electronic states of monolayers at thermal energies corresponding to temperatures of 298 K and (B) 473 K. Inverse grayscale colour bar describes the normalized valence electron density with white(black) denoting maximum(minimum) values. (C) Unfolded electronic band structure of pristine and N-vacant monolayer MoSi$_2$N$_4$(MoN) supercell, with the superimposed projected density of states (PDOS) of both cases. (D) Plane-average charge density of MoSi$_2$N$_4$(MoN) monolayer with N-site vacancy. Blue(Cyan) colour represents electron depletion(accumulation) region, respectively. (E) Band-decomposed charge density of N-site vacant MoSi$_2$N$_4$(MoN) monolayer at thermal energies corresponding to temperatures of 298 K and 473 K.}
\end{figure*}

The electronic band structures and the atom projected density of states (PDOS) of the six monolayers are shown in Figure \ref{Fig3}. 
All monolayers are metallic. The states at the vicinity of the Fermi level (E$\mathrm{_F}$) are dominantly contributed by the transition metal atoms with some minor contributions from the N atoms. 
To gain insight on the bonding characteristic and the charge distribution among the atoms, the electron localization function (ELF) is calculated and shown in the upper panels of Figure \ref{Fig4}(A-F) for each monolayer. ELF values of 1 (red) and 0 (blue) corresponds to perfect valence electron localization and delocalization, respectively. The valence electron distribution is highly localised around N but less around the transition metals Mo or W, suggesting the ionic nature of the bonding within the inner MN. Additionally, there are almost no valence electrons localised around Si as compared to Ge due to the higher electronegativity difference between Si and N. This indicates that the Si-N bonds are more ionic than Ge-N bonds, contributing to the higher elastic moduli of Si-based monolayers than their Ge-based counterparts [see Table \ref{Elastic}]. This behavior is also evident in the Bader charge analysis, which shows a higher amount of electron gain by the outer N atoms in the Si-based monolayers (2.2 e) as compared to the Ge-based monolayers (1.4 e) [see Table \ref{Electronic}]. 
The electronic states from Si (Ge) atoms are almost absent around the Fermi level. 

We further calculate the plane-averaged electrostatic potential of the six monolayers [see bottom panels of Figure \ref{Fig4}(A-F)]. For the \emph{heterolayer} MoSi$_2$N$_4$(WN) and MoGe$_2$N$_4$(WN) monolayers, the absence of out-of-plane mirror symmetry (arising from the MoN-WN inner core layer combination) results in an asymmetrical electron charge distribution along the out-of-plane direction. This asymmetrical potential creates a built-in electric dipole ($\Delta \Phi$) which is defined as the difference of the vacuum levels at both sides of the \emph{heterolayered} monolayers. 
Both MoSi$_2$N$_4$(WN) and MoGe$_2$N$_4$(WN) monolayers are almost identical with $\Delta \Phi = 0.32$ eV since $\Delta V$ is a direct consequence of the Janus-like MoN-WN combination in the inner core layers.

\subsection{Sublayer protection mechanism of conduction states and defect-induced `punch through' effect} 

We now investigate the spatial distribution of the electronic states in the monolayers within an energy window of $\pm k_BT$ around the Fermi level, $\varepsilon_F$. Such energy window can be regarded as a \emph{transport window} since the states residing within such energy window dominate the carrier conduction at temperature $T$. We consider two temperature of for $T = 298$ K ($k_BT = 0.026$ eV) and $T = 473$ K ($k_BT = 0.041$ eV). The band-decomposed charge density plots of the transport window for all six monolayers are shown in Figure \ref{Fig5}(A, B). 
The charge density of each monolayer is normalized in the grayscale spectrum, with 1 and 0 corresponds to the maximum and minimum charge density, respectively. 

The electronic states within the $\varepsilon_F \pm k_BT$ energy window are contributed dominantly by the Mo or W atoms and the the inner N atoms [see Figure \ref{Fig5}(A)]. 
The localization of the electronic states around $\varepsilon_F$ at the vicinity of the transition metal atoms is also found in some monolayer metallic-TMDCs, such as 1T-MoS$_2$, 1T(1T')-WS$_2$, 1T(1T')-MoTe$_2$ and 1T(1T')-WTe$_2$ \cite{Huang2023_MoS2, Sun2016, Moradpur2023_WS2, Huang2016}, but the presence of AN outer layers in MA$_2$N$_4$(MN) provides an intriguing \emph{sublayer protection mechanism} in which the conduction states around the Fermi level are spatially insulated from the outer environment by the outer AN sublayers -- a feature similar to the case of semiconducting MoSi$_2$N$_4$ and WSi$_2$N$_4$ where the conduction band edges states are spatially protected by the outer SiN layer \cite{Wang2021, wu2022prediction}. Carrier conduction mediated by states around $\varepsilon_F$ are thus expected to be robust against moderate external perturbations. 
The sublayer protection of the Fermi level states suggests the potential of MA$_2$N$_4$(MN) as a 2D metal for interconnect application in which the outer AN layer serves as a \emph{native barrier layer}.

To understand how defects may affect the electronic structures of MoA$_2$N$_4$(MN) monolayers, we construct a 4 $\times$ 4 supercell of MoSi$_2$N$_4$(MoN) with an N vacancy located at the outer SiN sublayer.  Figure \ref{Fig5}(C) shows the band structure of the supercell for the pristine and defective cases, with the superimposed PDOS displayed at the right. Non-dispersive `flat band' is formed around the Fermi level, which corresponds to the localized states created by the N vacancy defect. 
The PDOS plot shows that the total density of states in this narrow energy window is significantly higher than that of the pristine supercell, with the appearance of peaks dominantly contributed from Si and the delocalised states. The existence of Si dangling bonds at the N vacancy site create new states around the Si atom, and also generate substantial delocalised states around the N vacancy [see PDOS in Figure \ref{Fig5}(C)]. To further understand the charge distribution before and after defect formation, we calculate the plane-averaged differential charge density ($\Delta\rho$) [see Figure \ref{Fig5}(D)], defined as $\Delta\rho = \rho_{defective} - \rho_{pristine}$ where $\rho_{defective}$ and $\rho_{pristine}$ is the charge density of the defective and pristine supercell, respectively. The pronounced negative peak in the differential charge density profile suggests the generation of `punch through' states that connects the inner conduction channel to the external environment.
Such `punch through' states can be more clearly observed in the band-decomposed charge density of the defective supercell in Figure \ref{Fig5}(E). 
Importantly, the conduction states contributed by the inner Mo atoms remain largely unaltered by the newly formed electronic states induced by the N-vacant site. These aspects suggest a surface defect activation mechanism that forms conduction channel that bridges carrier conduction from the inner core layers to the outer environment. Interestingly, this behavior is akin to the `vias' structure of a metal interconnect through which the metal interconnect makes a vertical electrical contact to the source/drain of a transistor. 
MoSi$_2$N$_4$(MoN) is thus a potential candidate monolayer for interconnect applications due to three aspects: (i) the transition metal atoms in a 2D monolayer is resilient against electromigration effect \cite{hau2004introduction} -- a critical issue that plagues Cu interconnects \cite{tu2003recent, shen2023electromigration}; (ii) the conduction states residing in the inner core layers are protected natively by the SiN outer nitride sublayers, thus circumventing the need of a barrier layer such as the Ta/TaN barrier layer used in Cu-based interconnect \cite{li2020recent}; and (iii) vias connection can be activated by defect engineering.

\begin{figure*}
\centering
\includegraphics[width=1\textwidth]{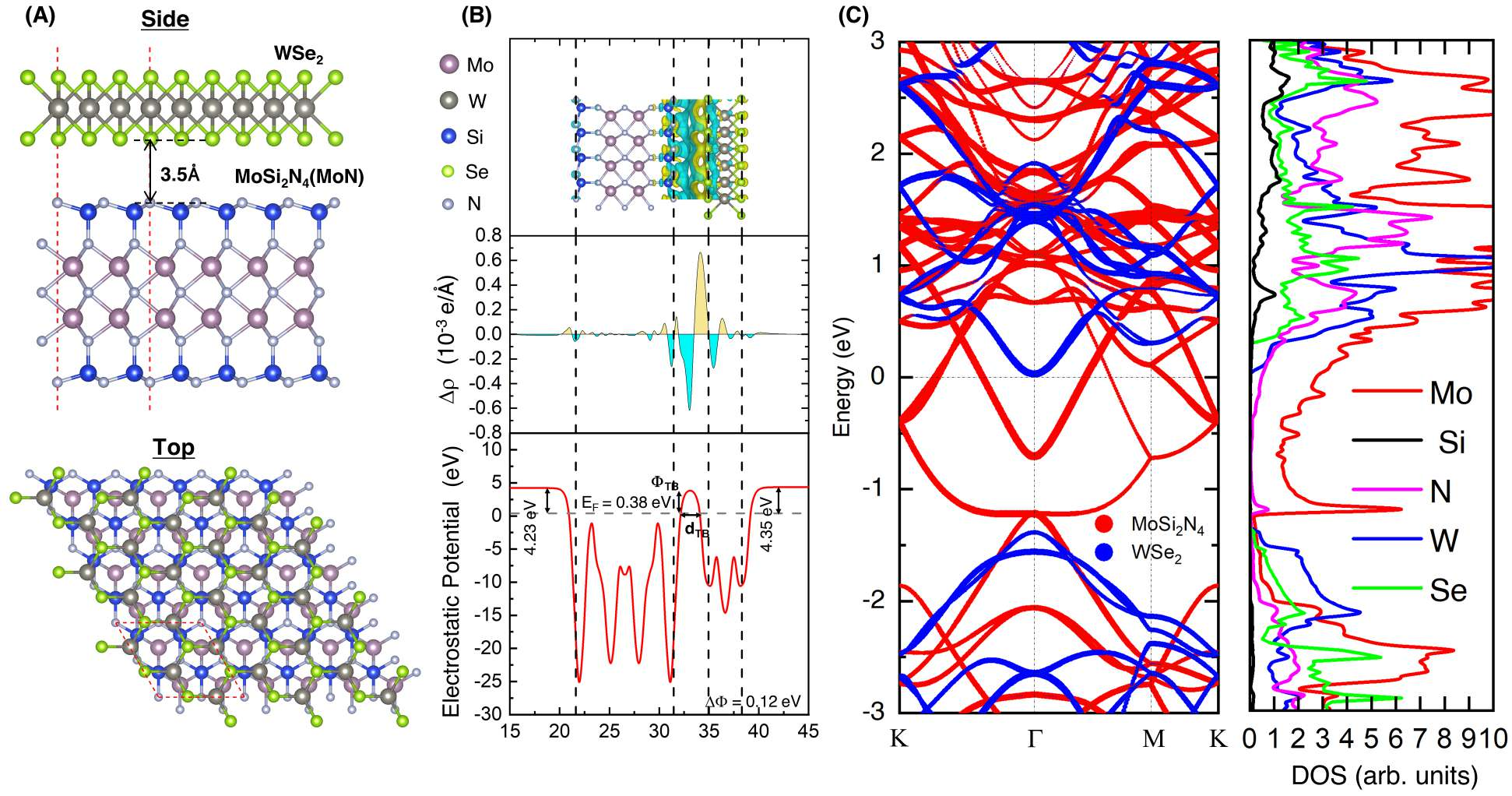}
\caption{\label{Fig6} \textbf{Lattice structure and electronic profiles of MoSi$_2$N$_4$(MoN)/WSe$_2$ vdWH.} (A) Side and top view of the lattice structure, showing the equilibrium interlayer distance and relative stacking order. Red dotted lines denotes the unit cell of the vdWH. (B) Plane-averaged differential charge density (upper and middle panel) and electrostatic potential profile (bottom panel) of the vdWH. Yellow(cyan)-shaded region represents electron accumulation(depletion). (C) Electronic band structure and projected density of states to atoms.}
\end{figure*}

\subsection{MoSi$_2$N$_4$(MoN) as a metal contact to 2D semiconductor}

We now investigate MoSi$_2$N$_4$(MoN) as a metallic electrode to 2D semicondcutor WSe$_2$ -- a widely studied 2D semiconductor for transistor applications \cite{Zhang2023_transistor, Li2023_transistor, Zhu2021, Li2022_transistor}. MoSi$_2$N$_4$(MoN)/WSe$_2$ contact is constructed with QuantumATK, using supercell consisting of  2 $\times$ 2 MoSi$_2$N$_4$(MoN) unit cells and $\sqrt{3} \times \sqrt{3}$ WSe$_2$ unit cells to take into account their lattice mismatch, with an equal strain of 1.25\% is applied to both monolayers. The contact heterostructure is geometrically optimized to [see Table \ref{Elastic} for the lattice constants]. The interlayer istance of 3.50 $\mathrm{\AA}$ is comparable to other 2D/2D contact heterostructures \cite{Pei2023}. 
MoSi$_2$N$_4$(MoN)/WSe$_2$ retains the hexagonal symmetry of MoSi$_2$N$_4$(MoN), and the mechanical stability follows the same Born-Huang criteria as that of standalone MA$_2$N$_4$(M'N) monolayers. The MoSi$_2$N$_4$(MoN)/WSe$_2$ is mechanically stable and the 3D elastic moduli are generally less than that obtained for the six MA$_2$N$_4$(MN) monolayers despite having a larger $Y^{2D}$ [see Table \ref{Elastic}].

\begin{figure*}
\centering
\includegraphics[width=0.85\textwidth]{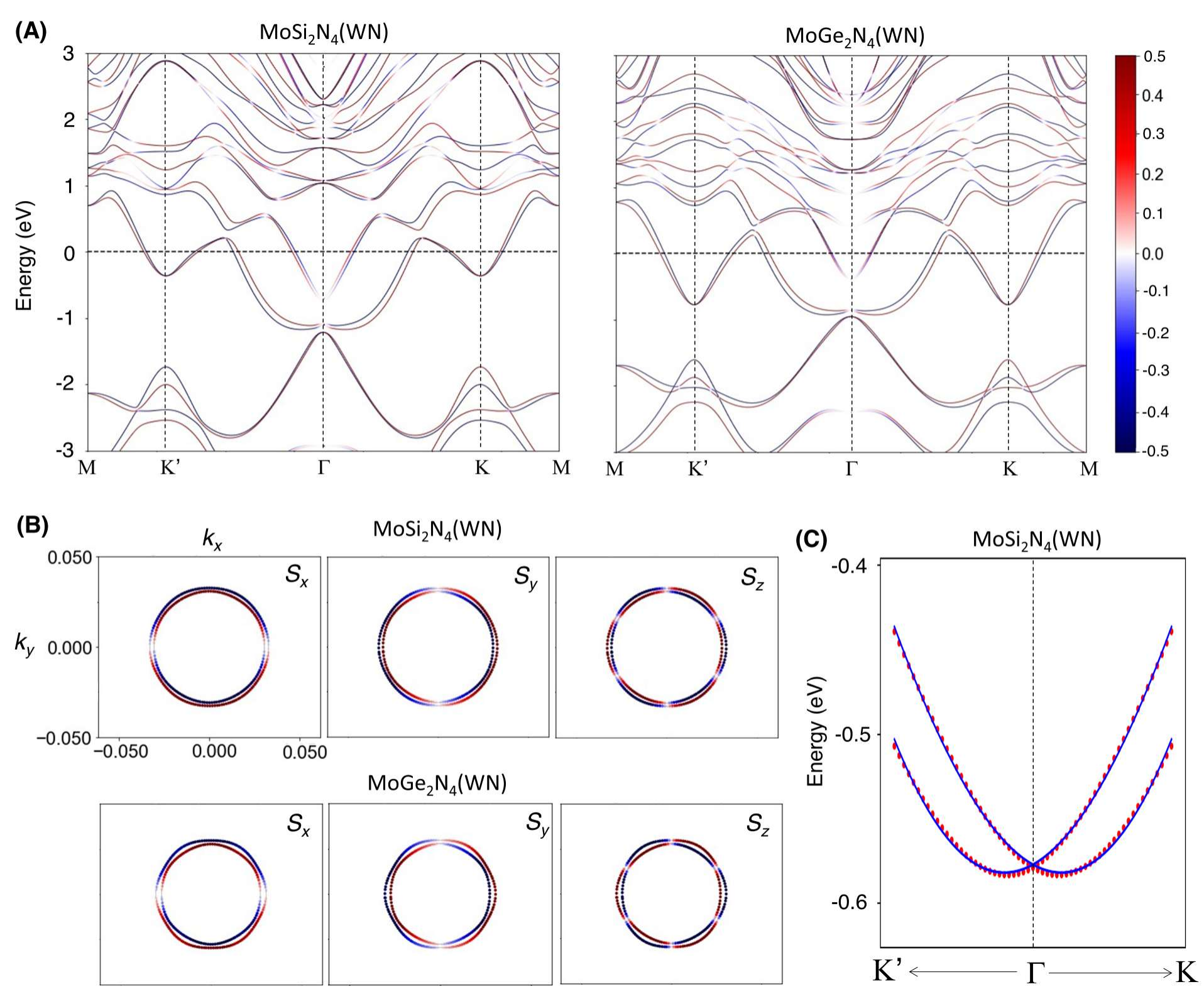}
\caption{\label{Fig7} \textbf{Electronic band structure with spin-orbit coupling and spin texture of \emph{heterolayered} monolayers.} (A) Electronic band structure of MoSi$_2$N$_4$(WN) (Left) and MoGe$_2$N$_4$(WN) (Right). Colour bar describes the $S_Z$ spin component with red(blue) denoting spin-up(spin-down). (B) Spin texture of MoSi$_2$N$_4$(WN) (top panels) at -0.5 eV and MoGe$_2$N$_4$(WN) (bottom panels) at -0.3 eV, projected along $S_X$, $S_Y$, $S_Z$ spin components. (C) Rashba-type spin-split bands of MoSi$_2$N$_4$(WN) in the vicinity of the $\Gamma$ point, where blue circles denote DFT obtained results and red line denotes our fitting model. }
\end{figure*}

The top and side views of the fully-relaxed lattice structure, the electrostatic potential profile with the plane-averaged differential charge density,  and the electronic band structures as well as the PDOS are shown in Figure \ref{Fig6}(A), \ref{Fig6}(B), \ref{Fig6}(C), respectively. Interface charge redistribution occurs upon forming the MoSi$_2$N$_4$(MoN)/WSe$_2$ contact, electrons are transferred from MoSi$_2$N$_4$(MoN) to WSe$_2$ [see top and middle panel of Figure \ref{Fig6}(B)] which establishes a built-in electric dipole as evident from the energy difference in the vacuum levels between the two sides of the contact [see bottom panel of Figure \ref{Fig7}(B)]. An ultralow Schottky barrier [see Figure \ref{Fig6}(C)] of 0.03 eV is formed. Such value is comparable to the thermal energy at room temperature (i.e. $k_BT \approx 0.26$ eV), thus suggesting the quasi-Ohmic contact of  MoSi$_2$N$_4$(MoN)/WSe$_2$ for devices operating at room temperatures \cite{Cao2021}. The electrostatic barrier formed at the vdW gap between the monolayers may impede charge injection efficiency [see bottom panel of Figure \ref{Fig6}(C)], which can be quantified by the tunnelling specific resistivity ($\rho_{t}$) through Simmon’s tunneling diode model \cite{Simmons1963, Cao2022}:
\begin{subequations}
\begin{equation}\label{tunneling_probability}
\mathcal{T}(\Phi_{TB},d_{TB})=\exp\left(\frac{-2d_{TB}\sqrt{2m_{e}\Phi_{TB}}}{\hbar}\right)
\end{equation}
\begin{equation}\label{tunneling_resistivity_intermediate}
\rho_{t}(\Phi_{TB},d_{TB})=\frac{\pi^{2}\hbar d_{TB}^{2}}{e^{2}\left(\frac{\sqrt{2m_{e}\Phi_{TB}}}{\hbar}-1\right)}\mathcal{T}^{-1}
\end{equation}
\end{subequations}
where $\hbar$, e, m$_e$ is the reduced Planck constant, charge magnitude and mass of the free electron respectively. Here the values of the tunnelling barrier height ($\Phi_{TB}$) and width ($d_{TB}$) are extracted from the plane-averaged electrostatic potential profile [see bottom panel of Figure \ref{Fig6}(B)]. 
The extracted values of $\Phi_{TB}$ and $d_{TB}$ from the electrostatic potential profile are 3.43 eV and 1.93 $\mathrm{\AA}$, respectively. 
The $\rho _t$ is calculated as 2.85 $\times$ 10$^{-9}$ $\Omega$ cm$^{2}$  which are in the same order of magnitude as several previously studied MoSi$_2$N$_4$-based contacts \cite{Wang2021,Tho2022,Liang,Zhang2023_MXene} and the recently reported (Bi, Sb)/WSe$_2$ \cite{Su2023_semimetal}.

\subsection{Spin-orbit coupling effect} 

The presence of $d$-orbital electrons in heavy transition metal atoms that can couple strongly with the intrinsic electric field of structures without inversion symmetry \cite{Zhou2021_SOC}, resulting in Rashba spin-orbit coupling (RSOC) effect. We include the spin-orbit coupling (SOC) effect for the heterolayered MoSi$_2$N$_4$(WN) and MoGe$_2$N$_4$(WN) [see Figure \ref{Fig7}(A)] due to their additional mirror symmetry breaking that can further increase the RSOC strength. Time reversal symmetry imposes the condition that the spin-split states connecting $\Gamma$ to $K$ and from $\Gamma$ to $K’$ must be of opposite signs, which is observed from the value of the colour map [0.5 ($-0.5$) denotes  full spin-up (spin-down)] when the states are projected on the out-of-plane spin component $S_z$. The spin-splitting effect in the heterolayered monolayers can be projected onto the full reciprocal space, which enables us to plot the in-plane spin texture at the constant energy of $-0.5$ eV and $-0.3$ eV near the $\Gamma$ point for MoSi$_2$N$_4$(WN) and MoSi$_2$N$_4$(WN), respectively. All three expectation values of the spin components (S$_x$, S$_y$, S$_z$) are non-zero when they are mapped out on the 2D reciprocal space ($k_{x}, k_{y}$) centered at the $\Gamma$ point. The spin texture rings calculated at a constant energy [see Figure \ref{Fig7}(B)] clearly reveal the RSOC effect. The in-plane (S$_x$, S$_y$) spin components rotate in opposite directions for the inner and outer rings, which is the characteristic of the in-plane RSOC caused by the broken mirror symmetry. The RSOC effect at the $\Gamma$ point in MoSi$_2$N$_4$(WN) is more pronounced compared to that in MoGe$_2$N$_4$(WN), which is due to the larger potential energy difference between W and Si compared to between W and Ge [see Figure \ref{Fig4}(C, F)], thus creating a stronger electric field that couples to the $d$-orbital electrons of the W atom \cite{Wu2022_SOC}.

We further construct a non-linear Rashba model for MoSi$_2$N$_4$(WN) that is beyond the simple linear form $H_{R}(k)=\alpha_{R}(k \times \hat{z})\cdot \sigma$, where $\alpha_{R}$ is the coupling strength, $\sigma$ is the vector of Pauli matrices ($\sigma_x$, $\sigma_y$, $\sigma_z$), and $k$ is the wave vector of the electron. To describe the non-zero S$_z$ component, the $k \cdot p$ model is expanded at $\Gamma$ to include higher order terms in the expansion. The two-fold degeneracy at $\Gamma$ corresponds to the $\Gamma_4$ double-value irreducible representation of point-group $C_{3v}$. Using these states as the basis, the two generators of $C_{3v}$ ($C_3$, $M_{xz}$) and the time reversal operator (\( \mathcal{T} \)) are represented as \cite{Guo2023}:
\begin{equation}\label{generators}
C_3 = e^{-i\pi \sigma_{z}/3}, \enspace M_{xz}=-i\sigma_{y}, \enspace \mathcal{T}=-i\sigma_{y}\mathcal{K}
\end{equation}
where \( \mathcal{K} \) is the complex conjugation operator. The effective $k \cdot p$ Hamiltonian ($H_{eff}(k)$) is constrained by Equation \ref{generators} to follow:
\begin{subequations}
\begin{equation}\label{C3}
C_{3}H_{eff}(k)C_{3}^{-1} = H_{eff}(R_{3}k)
\end{equation}
\begin{equation}\label{M_xz}
M_{xz}H_{eff}(k)M_{xz}^{-1} = H_{eff}(k_{x}, -k_{y}) 
\end{equation}
\begin{equation}\label{T}
\mathcal{T}H_{eff}(k)\mathcal{T}^{-1} = H_{eff}(-k) 
\end{equation}
\end{subequations}
where $R_{3}$ is a 120$^{\circ}$ rotation applied to the wave vector. By expanding $H_{eff}(k)$ to the $k^{4}$ order, we obtain the following form:
\begin{equation}\label{C3}
 \begin{aligned}
H_{eff}(k) = c_{1}(k_{x}\sigma_{y}-k_{y}\sigma_{x}) + c_{2}k^{2}\sigma_{0} + c_{3}(k_{x}\sigma_{y}-k_{y}\sigma_{x})k^{2} \\ + c_{4}(k_{y}^{3}-3k_{y}k_{x}^{2})\sigma_{z} + c_{5}k^{4}\sigma_{0}
 \end{aligned}
\end{equation}
where $\sigma_{0}$ is the 2 $\times$ 2 identity matrix. Using this model, we obtained the parameter values of $c_{1} = 4.30 \ \mathrm{eV \ \AA}$, $c_{2} = 26.47 \ \mathrm{eV \ \AA^{2}}$, $c_{3} = -0.64 \ \mathrm{eV \ \AA^{3}}$, $c_{4} = 33.84 \ \mathrm{eV \ \AA^{4}}$ and $c_{5} = -46.93 \ \mathrm{eV \ \AA^{5}}$ for MoSi$_2$N$_4$(WN), which is in agreement with the result obtained from DFT [see Figure \ref{Fig7}(C)]. We note that by considering the $k$-linear term of $H_{R}(k)$ form, the Rashba SOC strength parameter of MoSi$_2$N$_4$(WN) can be written as $\alpha_R = c_{1}$, which allows us to benchmark the extent of spin splitting with several other materials as shown in Table \ref{Rashba}. The much higher $\alpha_{R}$ value obtained in our work compared to the bulk metals in Table \ref{Rashba} can host several advantages over them, for example it circumvents the need to deposit a thin film of metal onto the bulk substrate in order to increase the Rashba splitting strength of its surface electrons. Therefore, the high $\alpha_{R}$ values found in 2D metallic systems can open up new opportunities in the area of spintronics.

\begin{table}[t]
{
\caption{\label{Rashba} Rashba coefficient ($\alpha_{R}$) under the linear Rashba Hamiltonian form ($H_{R}(k)$) for the result of this work and several other materials reported in literature.}
\begin{tabular}{>{\centering\arraybackslash}m{2.5cm}>{\centering\arraybackslash}m{2cm}>{\centering\arraybackslash}m{2cm}}
\hline
\hline

Material & $\alpha_R$ (eV $\mathrm{\AA}$) & Reference \\ \hline \\

MoSi$_2$N$_4$(WN) & 4.3 & Current Work \\ \\ \hline \\

\emph{Metal Surface States} & & \\ \\

Pb/Ge(111) & 0.24 & \cite{Yaji2010} \\ \\

Ag(111) & 0.03 & \cite{Popovic2005} \\ \\

Au(111) & 0.33 & \cite{Lashell1996} \\ \\

Bi(111) & 0.55 & \cite{Koroteev2004} \\ \\ \hline \\

\emph{2D Semiconductor} & & \\ \\

SbP & 1.103 & \cite{Guo2023_rashba} \\
BiP & 2.454 & \cite{Guo2023_rashba} \\
BiAs & 4.359 & \cite{Guo2023_rashba} \\
BiSb & 7.113 & \cite{Guo2023_rashba} \\ \\

GaSeCl & 1.2 & \cite{Sasmito2021_Rashba} \\
GaTeI & 1.9 & \cite{Sasmito2021_Rashba} \\
GaTeBr & 2.4 & \cite{Sasmito2021_Rashba} \\
GaTeCl & 2.65 & \cite{Sasmito2021_Rashba} \\ \\

MoSSe & 0.077 & \cite{Hu2018_Rashba} \\
MoSTe & 0.148 & \cite{Hu2018_Rashba} \\
MoSeTe & 0.487 & \cite{Hu2018_Rashba} \\
WSSe & 0.157 & \cite{Hu2018_Rashba} \\
WSTe & 0.324 & \cite{Hu2018_Rashba} \\
WSeTe & 0.524 & \cite{Hu2018_Rashba} \\ \\

Si$_2$SeTe & 0.022 & \cite{Guo2023} \\ \\

\hline
\hline
\end{tabular}
}
\end{table}

\section{\label{sec:Conclusion}Conclusion}

In summary, we investigated a series of stable MA$_2$N$_4$(M'N) monolayers based on first-principles calculations. The six monolayers are metallic and the heterolayed monolayers exhibit sizable RSOC effect. Intriguingly, the conduction states around E$_F$ is localised within the inner core layers, and the outer passivating nitride sublayers provide a built-in sublayer protection mechanism that spatially insulates the conduction states from external environment. 
Creating N vacancy defects at the outer sublayer introduces new electronic states at the E$_F$, which provides a `punch through' pathway to connect the conduction states in the inner core layers with the external environment, thus potentially providing a `vias'-like electrical contact structure. Furthermore, MoSi$_2$N$_4$(MoN)/WSe$_2$ forms a quasi-Ohmic contact beneficial for devices operating at room temperatures. Finally, we note that MoSi$_2$N$_4$(MoN) is composed exclusively of elements posing low-risk to environment and human health, thus suggesting its potential compatibility for developing next-generation sustainable semiconductor device technology \cite{https://doi.org/10.1002/adfm.202308679}. Our study provides a hindsight on the potential of ultrathick MA$_2$Z$_4$(M'Z) for technological applications, hinting that rich and applicable physical properties may be uncovered from the future studies of the broader MA$_2$Z$_4$(M'Z)$_n$ family.

\section*{\label{sec:Acknowledgement}Acknowledgement}
This work is funded by the Singapore Ministry of Education (MOE) Academic Research Fund (AcRF) Tier 2 Grant (MOE-T2EP50221-0019) and SUTD-ZJU IDEA Thematic Research Grant ().

\section*{\label{sec:Declaration}Author Declarations}
\subsection*{Conflicts of Interest}
The authors have no conflict to disclose. 

\subsection*{Author Contributions}
\textbf{Che Chen Tho}: Data curation (lead); Formal analysis (lead);
Investigation (equal); Writing – original draft (equal); Visualization (lead); Writing – review \& editing (supporting);
\textbf{Xukun Feng}: Investigation (equal); Formal analysis (equal); Writing – original draft (supporting); 
Investigation (equal); Writing – original draft (equal); Visualization (lead); Writing – review \& editing (equal);
\textbf{Liemao Cao}: Writing – review \& editing (supporting);
\textbf{Zhuoling Jiang}: Writing – review \& editing (supporting);
\textbf{Guangzhao Wang}: Conceptualization (supporting); Data curation (supporting); Writing – original draft (supporting); Writing – review \& editing (supporting);
\textbf{Chit Siong Lau}: Writing – review \& editing (supporting);
\textbf{San-Dong Guo}: Formal analysis (supporting); Writing – review \& editing (supporting);
\textbf{Yee Sin Ang}: Conceptualization (lead); Investigation (equal); Formal analysis (supporting); Supervision (lead); Funding acquisition (lead); Visualization (supporting); Writing – original draft (equal); Writing – review \& editing (lead).

\section*{Data Availability Statement}

Data sharing is not applicable to this article as no new data were
created in this study.

\providecommand{\noopsort}[1]{}\providecommand{\singleletter}[1]{#1}%

\end{document}